\documentclass[reprint, aps, prd, superscriptaddress, nofootinbib]{revtex4-2}

\usepackage{graphicx}
\usepackage{amsmath}
\usepackage{amssymb}
\usepackage{orcidlink}
\usepackage{ulem}

\usepackage{xcolor, soul}

\newcommand{\e}{\mathrm{e}}    
\newcommand{\ddd}{\mathrm{d}}  
\newcommand{\du}{\mathrm{d}}   
\newcommand{\iu}{\mathrm{i}}   
\newcounter{count}
\newcommand*{\defeq}{\mathrel{\vcenter{\baselineskip0.5ex \lineskiplimit0pt
                     \hbox{\scriptsize.}\hbox{\scriptsize.}}}%
                     =}

\newcommand{\myRe}{\mathsf{Re}}

\graphicspath{{./figures/}}

\newcommand{\jena}{Institute for Theoretical Physics, University of Jena, Fröbelstieg 1, 07743, Jena, Germany}
\newcommand{\damtp}{Department of Applied Mathematics and Theoretical Physics, Centre for Mathematical Sciences, University of Cambridge, Wilberforce Road, Cambridge CB3 0WA, United Kingdom}
\newcommand{\bham}{Institute for Gravitational Wave Astronomy \& School of Physics and Astronomy, University of Birmingham, Edgbaston, Birmingham B15 2TT, UK}
\newcommand{\kavli}{Kavli Institute for Cosmology Cambridge, Madingley Road CB3 0HA, Cambridge, UK}
\newcommand{\jhu}{Department of Physics and Astronomy, Johns Hopkins University, 3400 N Charles Street, Baltimore, Maryland 21218, USA}
\newcommand{\caltech}{Theoretical Astrophysics 350-17, California Institute of Technology, 1200 E California Boulevard, Pasadena, California 91125, USA}

\begin{document}

\normalem

\title{Stochastic gravitational wave background from supernovae in massive scalar-tensor gravity}

\author{Roxana Rosca-Mead \orcidlink{0000-0001-5666-1033}}
\email{roxana.rosca-mead@uni-jena.de}
\affiliation{\jena}

\author{Michalis Agathos \orcidlink{0000-0002-9072-1121}}
\email{magathos@damtp.cam.ac.uk}
\affiliation{\kavli}
\affiliation{\damtp}

\author{Christopher J. Moore \orcidlink{0000-0002-2527-0213}}
\email{cmoore@star.sr.bham.ac.uk}
\affiliation{\bham}

\author{Ulrich Sperhake \orcidlink{0000-0002-3134-7088}}
\email{us248@maths.cam.ac.uk}
\affiliation{\damtp}
\affiliation{\jhu}
\affiliation{\caltech}

\date{\today}

\begin{abstract}
  In massive scalar-tensor gravity, core-collapse supernovae are strong sources of scalar-polarized gravitational waves.
  These can be detectable out to large distances.
  The dispersive nature of the propagation of waves in the massive scalar field implies that the gravitational wave signals are long-lived and many such signals can overlap to form a stochastic background. Using different models for the population of supernova events in the nearby universe, we compute predictions for the energy density in the stochastic scalar-polarized gravitational wave background from core-collapse events in massive scalar-tensor gravity for theory parameters that facilitate strong scalarization.
  The resulting energy density is below the current constraints on a Gaussian stochastic gravitational wave background but large enough to be detectable with the current generation of detectors when they reach design sensitivity, indicating that it will soon be possible to place new constraints on the parameter space of massive scalar-tensor gravity.
\end{abstract}

\maketitle

\section{Introduction}\label{sec:introduction}

Gravitational waves (GWs) offer new opportunities for studying astrophysics and fundamental physics.
Before the first detection of a GW on 14 September 2015 by the two LIGO interferometers~\cite{LIGOScientific:2016aoc}, tests of general relativity (GR) have been largely limited to the weak-field regime.
The LIGO~\cite{LIGOScientific:2014pky} and Virgo~\cite{VIRGO:2014yos} observatories provide the opportunity to extend these to the strong and dynamical regimes: examples include the use of the first detection to test gravity~\cite{2016PhRvL.116v1101A}; the first GW polarization tests using simultaneous observations from three interferometers~\cite{PhysRevLett.119.141101}; setting constraints on both the properties of nuclear matter~\cite{2018arXiv180511581T} and on the speed of GWs~\cite{2041-8205-848-2-L13} from the first binary neutron star merger; and the suite of tests performed with the most recent GW catalog of detected signals from compact binary coalescence events~\cite{2021arXiv211206861T}.
GWs provide us with a new tool to use in examining possible beyond-GR theoretical frameworks.

Supernovae have long been considered as possible sources of GWs~\cite{PhysRevLett.17.1228} and,
in spite of the lack of positive detection as of now, they
remain a target for modern GW interferometers~\cite{2016PhRvD..94j2001A}.
Birkhoff's theorem guarantees that spherical symmetry implies no GWs (in GR).
Therefore, studying GW emission from core collapse requires multidimensional simulations. 
Notwithstanding the computational costs, it is now possible to obtain predictions for the GW signal from full $3+1$ dimensional simulations of core collapse (see, for example, Refs.~\cite{doi:10.1093/mnras/stx618, 2016ApJ...829L..14K, 0004-637X-851-1-62}). 
Using the results from such simulations, previous work~\cite{2005PhRvD..72h4001B,Crocker:2017agi,Finkel:2021zgf} has estimated the magnitude of the stochastic GW background (SGWB) due to supernovae; they found that the background is Gaussian for frequencies below $1\,\mathrm{Hz}$ and is small but may be relevant for future space-based GW missions, such as DECIGO~\cite{2021PTEP.2021eA105K}.

Scalar-tensor (ST) theories of gravity have long been considered as good candidates for modified gravity. 
Scalars arise naturally in high-dimensional theories, including string theory, and feature prominently in beyond-$\Lambda$CDM cosmologies, for example, as drivers of inflation or as candidates for dark energy.
Furthermore, ST theories have a well-posed Cauchy formulation~\cite{Faraoni:2004pi,Fujii:2003pa, 2015LNP...892....3S}.
ST theories date back to the pioneering work of Brans and Dicke~\cite{Fierz1956128,Jordan1959,PhysRev.124.925}.
More general Horndeski theories also give second order equations of motion~\cite{1974IJTP...10..363H}, but ST theories can mimic GR in weak fields, thereby passing all solar system tests, nevertheless exhibiting large deviations in the strong-field regime.
Spontaneous scalarization of neutron stars is such a possible signature of ST theories~\cite{PhysRevD.54.1474, PhysRevLett.70.2220}.
Because the scalar field is nonminimally coupled to the metric, gravity is now mediated by both tensor and scalar degrees of freedom.
GW emission can occur even in spherical symmetry in these theories, which in itself would be a fundamental deviation from GR\@.
Previous work has used simulations in $1+1$ dimensions of dynamical spontaneous scalarization during spherically symmetric core collapse to investigate possible GW signals in ST theories with a massless scalar~\cite{2000ApJ...533..392N,THESIS,0264-9381-33-13-135002}.

Through a conformal transformation of the physical metric $g_{\mu\nu}=\bar{g}_{\mu\nu}/F(\varphi)$, the theory can be written in the so-called Einstein frame where it resembles GR minimally coupled to a scalar field (see Sec.~\ref{subsec:action}).
This coupling function represents 1 degree of freedom of the theory.
A common choice for this function consists in a series
expansion of $\log F$ in the scalar field variable
$\varphi$.
Here, we employ such an expansion to second order,
\begin{align}\label{eq:coupling}
    \log F(\varphi)=-2\alpha_{0}\varphi-\beta_{0}\varphi^{2} \,.
\end{align}
This class of theories is sometimes referred to as Damour-Esposito Far{\`e}se (DEF) theory~\cite{PhysRevLett.70.2220}.

For a zero mass of the scalar,
the field $\varphi$ is long ranged ($\sim\!1/r$), and the parameters
$\alpha_0$ and $\beta_0$ of the theory are rather tightly
constrained by observations.
Specifically, observations by the Cassini mission
\cite{2003Natur.425..374B} are compatible with
{\it massless\/} ST theory only for
$\alpha_{0}<3.4\times 10^{-3}$, while the best constraints on the quadratic term come from timing of the binary pulsars PSR J1738+0333~\cite{2012MNRAS.423.3328F} and PSR J0348+0432~\cite{2013Sci...340..448A} which
rule out values $\beta_0 \lesssim -5$;
see also Ref.~\cite{2012MNRAS.423.3328F} and
Fig.~1 in Ref.~\cite{0264-9381-33-13-135002}.
For discussion of these constraints, see~\cite{6d97dcf5be1a4f88aed4ff8b4d1635b7} (in particular, Fig.~37).
This is problematic as it eliminates most of the parameter values for which spontaneous scalarization occurs.

A natural extension of such theories is to give the scalar field a mass, $\mu$.
Spontaneous scalarization still happens in massive ST theories and 
ST theories, with scalar field masses $\mu\gtrsim 10^{-16}$ eV are essentially unconstrained~\cite{PhysRevD.93.064005} due to the short-range (Yukawa-type $\sim\!\e^{-2\pi r\mu/\hbar}/r$) falloff of the scalar field~\cite{PhysRevD.85.064041,PhysRevD.93.064005,2017PhRvL.119t1103S}.
The authors have previously investigated the generation and propagation of GW signals from core collapse in massive ST theories~\cite{2017PhRvL.119t1103S}.
More specifically, we have found that these signals can be (i) loud, and hence detectable at a relatively high rate out to large distances, and (ii) of long duration due to the dispersive nature of the propagation of the massive GW mode.
This motivates our present work as the resulting GW signals overlap and form a SGWB.
Recent LIGO limits on SGWBs with alternative polarizations~\cite{PhysRevLett.120.201102} may help constrain these theories.

In this paper we continue our analysis into the detection possibilities offered by massive ST theories in the case of core collapse.
Using results from our previous papers, we combine waveforms obtained from different initial stars, at different points in time, to build a stochastic background.
In Sec.~\ref{sec:theory} we cover the formalism and numerics employed and the propagation of GWs over cosmological distances.
In Sec.~\ref{sec:results} we begin by reviewing the qualitative and quantitative dependence of the results on the scalar parameter $\beta_0$.
We further describe the six GW sets used in building the stochastic backgrounds and the potential for detection in LIGO-VIRGO data.
Our conclusions are given in Sec.~\ref{sec:discussion} and we
present more details of the calculations for the long-distance GW propagation in Appendix~\ref{app:A}.

\section{Theory}\label{sec:theory}
\begin{figure}[t]
    \centering
    \includegraphics[width=0.48\textwidth]{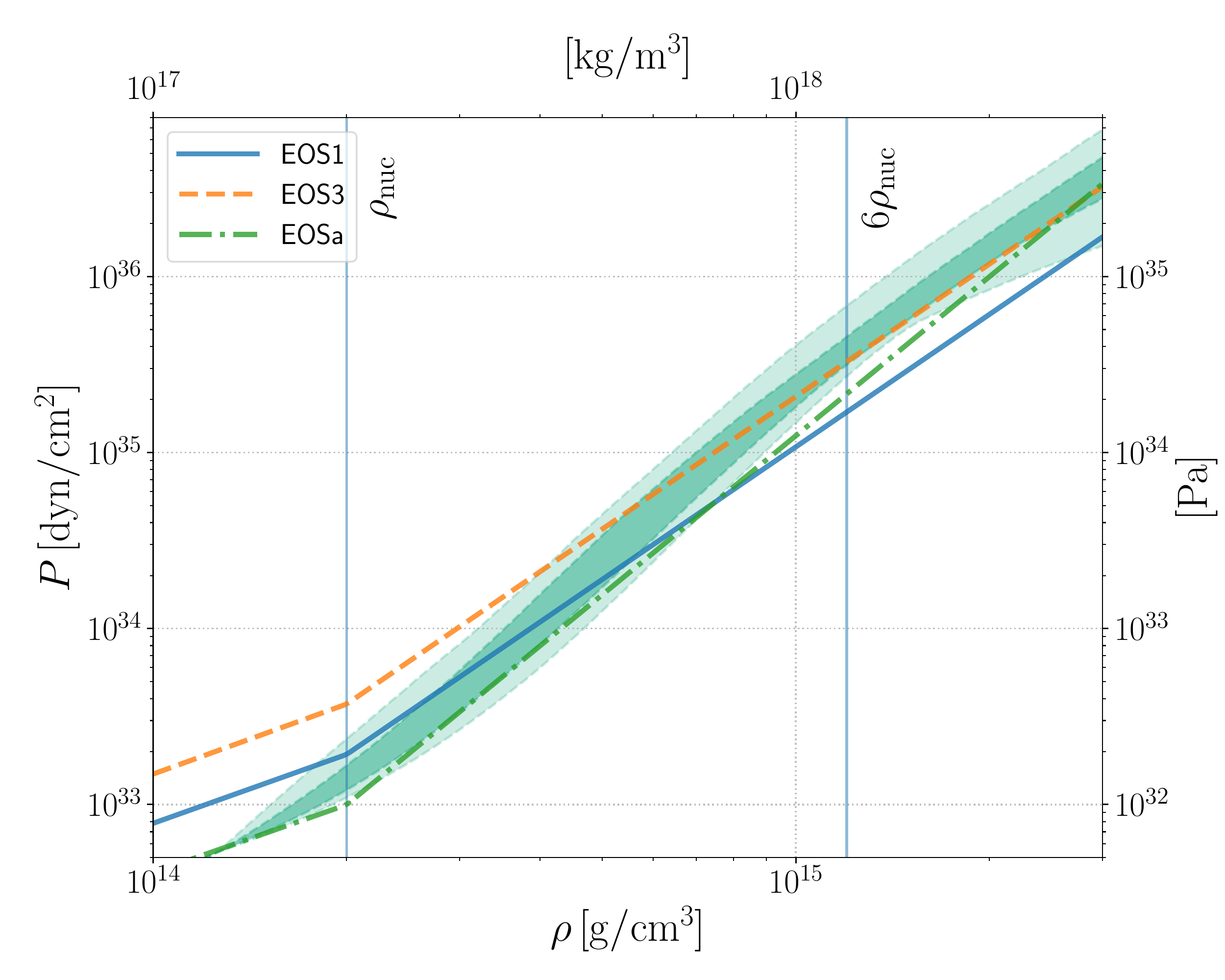}
    \caption{Pressure as a function of baryon density for the nonthermal component of the three EOS models used in this work. The values of the adiabatic exponents for the three models are given in Table \ref{tab:EOSparams}. The overlaid shaded regions correspond to the $90\%$ (light) and $50\%$ (dark)
    confidence bands given by the LVC analysis of GW170817 using the spectral parametrization for the EOS~\cite{2018arXiv180511581T}. Vertical lines indicate density values of one and six times the nuclear saturation density.
    }\label{fig:EOSfig}
\end{figure}

\subsection{ST action}
\label{subsec:action}
In the Einstein frame, the action can be written in the following form (natural units where $G = c = 1$ are used; however, factors of $\hbar$ are left explicit)
\begin{align}\label{eq:action}
    S\!=\!\int\ddd x^{4}\, \frac{\sqrt{-\bar{g}}}{16\pi} \big[ \bar{R} - 2\bar{g}^{\mu\nu}\partial_{\mu}\varphi\partial_{\nu}\varphi - 4V(\varphi) \big] \!+\! S_{m} \,,
\end{align}
where the matter action $S_m$ is taken to be that of a perfect (nuclear) fluid.
As discussed in Ref.~\cite{2017PhRvL.119t1103S}, this is a generic action for ST theories with a single nonminimally coupled scalar field. 
In Eq.~(\ref{eq:action}) $\varphi$ is the scalar field; $\bar{g}_{\mu\nu}$ is the conformal metric which is related to the physical metric by the coupling function in Eq.~(\ref{eq:coupling}) via $\bar{g}_{\mu\nu}=F(\varphi)g_{\mu\nu}$; $\bar{R}$ and $\bar{g}$ are the Ricci scalar and metric determinant associated with the conformal metric; and $V(\varphi)$ is the scalar field potential, which is taken to be a quadratic, $V(\varphi)=\hbar^{-2} \mu^{2}\varphi^{2}/2$, with scalar mass $\mu$.
The matter action $S_{m}$ is coupled to the physical metric $g_{\mu\nu}$ and is taken to describe a perfect fluid here.
The equations of motion for this theory are given by Eqs.~(3) in~\cite{2017PhRvL.119t1103S}.

\subsection{Equation of state}
It is necessary to further specify the equation of state (EOS) for the nuclear fluid. 
In this work, we employ hybrid EOSs that 
prescribe the pressure $P$ of the fluid as a function
of the mass density and the internal energy (or
temperature). More specifically, we use an EOS
first introduced in Ref.~\cite{1993A&A...268..360J}
that qualitatively captures in the form of a
{\it cold\/} pressure component the stiffening of matter
at nuclear densities and includes a {\it thermal\/}
term to model the response of shocked material. This EOS
is characterized by three parameters: two adiabatic
indices $\Gamma_1$ and $\Gamma_2$ for the cold
contribution at low and high densities, respectively,
as well as an
adiabatic index $\Gamma_{\rm th}$ for the thermal
pressure component $P_{\rm th}$. The full expressions, together
with a more detailed discussion, are given in
Sec.~3.1 of Ref.~\cite{0264-9381-33-13-135002}.
In the remainder of this work, we focus on three choices for this hybrid EOS given by the parameter combinations
listed in Table~\ref{tab:EOSparams}; the resulting cold pressure components are plotted in Fig.~\ref{fig:EOSfig}. 

\begin{table}
\caption{\label{tab:EOSparams}Parameter values describing the three EOSs used throughout the remainder of this paper.
} 
\begin{ruledtabular}
\begin{tabular}{llll}
        \phantom{} & $\Gamma_{1}$ & $\Gamma_{2}$ & $\Gamma_{\rm th}$ \\ \hline 
        EOS1 & 1.30 & 2.5 & 1.35 \\
        EOS3 & 1.32 & 2.5 & 1.35 \\
        EOSa & 1.28 & 3.0 & 1.50 
    \end{tabular}
\end{ruledtabular}
\end{table}

\subsection{Numerics}

We assume spherical symmetry and evolve the system of equations using a version of the open source \textsc{gr1d} code~\cite{2010CQGra..27k4103O} which is built to simulate stellar core collapse using finite differences and high-resolution shock capturing. The code was modified to include the scalar field in Ref.~\cite{0264-9381-33-13-135002} and the potential term was included later in Refs.~\cite{2020PhRvD.102d4010R,Rosca-Mead:2019seq}.
We use a uniform inner grid up to $r=40 $ km and a logarithmic one outside, up to $r=9\times10^5$ km.
The code shows between first and second order convergence with a discretization error of $4\%$ for the lowest resolution employed: $\delta r=250 $ m and $N=10000$ grid points for the inner grid
\cite{2017PhRvL.119t1103S}.

We initialize our data with spherically symmetric, nonrotating, stellar profiles from the catalog of Woosley and Heger~\cite{Woosley:2007as} who evolve Newtonian stars up to the point of iron core collapse.
The resulting progenitor configurations cover zero-age-main-sequence (ZAMS) masses ranging from 11 $M_\odot$ to 75 $M_\odot$ and three metallicities: solar ($\mathcal{Z}_\odot$), subsolar ($10^{-4} \mathcal{Z}_\odot$), and primordial (zero);
more details about these initial data can be found
in Ref.~\cite{Woosley:2007as} and Sec.~3.3 of Ref.~\cite{0264-9381-33-13-135002}.

\begin{figure*}[t]
    \centering
    \includegraphics[width=1.0\linewidth]{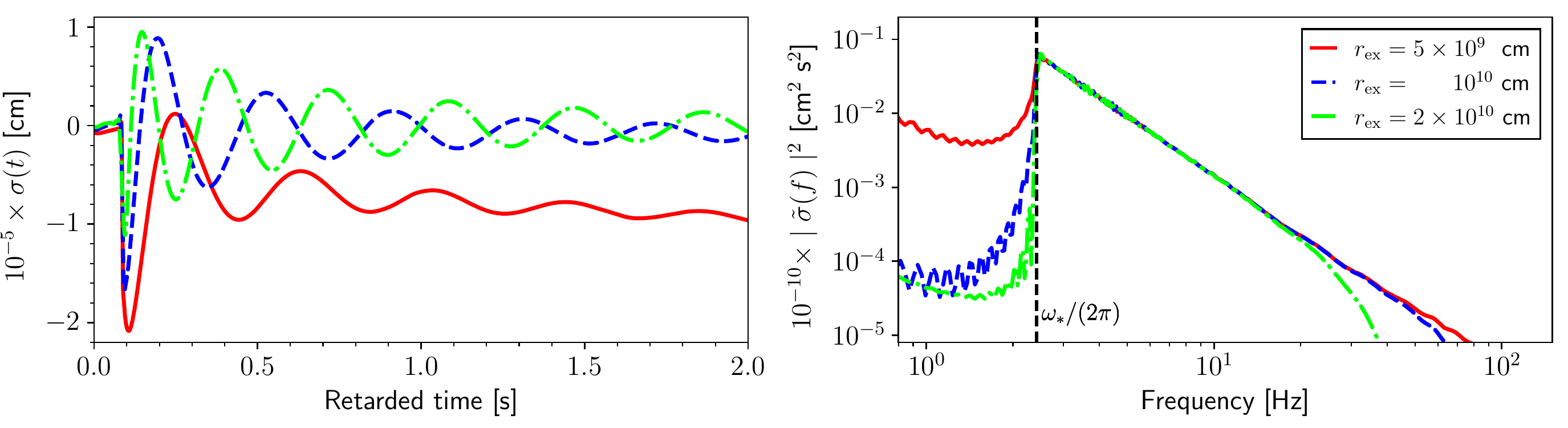}
    \caption{\label{fig:dispersive_nature}
        {\it Left panel\/}: example scalar signal $\sigma(t)=r\varphi(t)$ obtained from the core-collapse simulation of a $39 \, M_\odot$ star of $10^{-4}$ solar metallicity and EOS1 when $\mu=10^{-14}$ eV, $\alpha_0=10^{-2}$ and $\beta_0=-20$. The signal was extracted at several different radii ($5\times 10^9$ cm, $10^{10}$ cm and $2\times 10^{10}$ cm) and is plotted against retarded time.
        The drift observed in the case of $r_{\rm ex}=5\times 10^9$ cm (red line) is suppressed by the time the signal propagates outwards to the larger extraction radii; furthermore, as the signal travels outwards it becomes more oscillatory (see the discussion of dispersion and the explanation of the origin of the ``inverse chirp'' in~\cite{2017PhRvL.119t1103S}).
        {\it Right panel\/}: power spectrum of the waveforms plotted in the frequency domain. The low frequencies are exponentially suppressed when the signal propagates to large distances whereas the high frequencies propagate unimpeded.}
\end{figure*}

We can already see in the resulting numerical evolutions
the main effects of the wave propagation that reappear
in more dramatic form in the passage of the signal
across astrophysical distances. This is
illustrated in
Fig.~\ref{fig:dispersive_nature} where we show the
signal $\sigma\propto r\varphi$ [see Eq.~(\ref{eq:defsigma})] in the time and frequency domains
for a $39~M_{\odot}$ progenitor star with $10^{-4}$
solar metalicity, EOS1, and ST parameters $\mu=10^{-14}~{\rm eV}$,
$\alpha_0=10^{-2}$, $\beta_0=-20$ at different
radii. As the GW propagates outwards, it becomes
increasingly oscillatory, and low frequency contributions
below $\omega_*$ are further and further suppressed.
These features will be discussed in more detail in
the next subsection.

\subsection{Wave propagation}
\label{sec:waveprop}

For our calculation of the SGWB
generated by supernova events in the local universe,
we need to take into account the cosmological expansion
and its impact on the propagation of the scalar radiation.
For this purpose, we consider the spatially flat
Friedmann-Lema{\^\i}tre-Robertson-Walker (FLRW) metric,
\begin{equation}
  \du s^2 = -\du t^2 + a(t)^2(\du r^2+r^2\du \Omega^2)\,,
  \label{eq:ds2_FLRW}
\end{equation}
which is a good approximation for the background metric
in our cosmological neighborhood, i.e.~for distances
well below the curvature length scale of the universe.

The wave equation for a spherically symmetric
scalar field on this background is most conveniently
formulated in terms of conformal time $\eta$
and the radially rescaled scalar field $\sigma$ defined by
\begin{equation}
  \frac{\du \eta}{\du t} = \frac{1}{a}\,,
  ~~~~~
  \sigma = ar\varphi\,,
  \label{eq:defsigma}
\end{equation}
and it can be written as
\begin{equation}
  \partial_{\eta}^2 \sigma
  -\partial_r^2 \sigma
  -a^2 H^2(1-q)\sigma
  +\mu^2 a^2\sigma\
  = 0\,.
  \label{eq:wavetime}
\end{equation}
Here, $\sigma=\sigma(\eta,r)$ while $a$, $H$, and $q$
are functions of conformal time $\eta$ only and denote
the scale factor, the Hubble parameter $H=\dot{a}/a$, and
the deceleration parameter $q=-a\ddot{a}/\dot{a}^2$, with
$\dot{}=\du/\du t$, respectively. By comparing
Eq.~(\ref{eq:wavetime}) with its Minkowskian counterpart
[cf.~Eq.~(36) in Ref.~\cite{2020PhRvD.102d4010R}],
\begin{equation}
  \partial_t^2\sigma-\partial_r^2\sigma + \mu^2\sigma=0\,,
\end{equation}
we notice two key differences: the extra
term {\mbox{$-a^2 H^2(1-q)\sigma$}} and the additional
factor $a^2$ in the mass term. An analytic treatment
analogous to the Minkowski case is still possible, however,
if we treat the emission as instantaneous and the observation of the wave
signal as another instantaneous event relative to the timescale of
cosmological expansion\footnote{This is, of course,
a common assumption in most astrophysical observations where
the cosmological redshift of a source is treated as constant over the duration of the observation. In our case, the wave signals
are generated by a collapse lasting seconds and may
disperse into long signals over months or years, both clearly
well below the timescale of changes in the scale factor $a$.}. We can then treat $a$,
$H$, and $q$ as constants as we Fourier transform
Eq.~(\ref{eq:wavetime}) using the convention
\begin{equation}
  \tilde{f}(\omega)=\int_{-\infty}^{\infty}
  f(\eta)e^{\iu \omega \eta}\du \eta\,,
  ~~~
  f(\eta)=\frac{1}{2\pi}\int_{-\infty}^{\infty}
  \tilde{f}(\omega)e^{-\iu\omega \eta}\du \omega\,.
  \label{eq:FT}
\end{equation}
This results in the equation
\begin{eqnarray}
  \partial_r^2\tilde{\sigma}(\omega,r)
  &=&
  -(\omega^2 - \omega_*^2) \tilde{\sigma}(\omega,r)
  \,,~~\text{with}
  \nonumber
  \\[10pt]
  \omega_* &=& a\sqrt{\mu^2-H^2(1-q)}\,,
  \label{eq:waveFT1}
\end{eqnarray}
which is solved by
\begin{eqnarray}
  && \tilde{\sigma}(\omega,r)
  = \tilde{f}(\omega)e^{\iu k(r-r_{\rm e})}
  + \tilde{g}(\omega)e^{-\iu k(r-r_{\rm e})}\,,
  \nonumber
  \\[10pt]
  &&k = \sqrt{\omega^2-\omega_*^2}\,,
  \nonumber
  \\[10pt]
  &&\sigma(\eta,r) = \frac{1}{2\pi}\int_{-\infty}^{\infty}
  \tilde{f}(\omega)e^{\iu [k(r-r_{\rm e})-\omega\eta]}
  \nonumber \\[10pt]
  && \hspace{2.5cm}
  +\tilde{g}(\omega)e^{-\iu[k(r-r_{\rm e})+\omega \eta]}
  \du \omega
  \nonumber
  \\
  \label{eq:waveFTgensol}
\end{eqnarray}

The result for our specific case of a real signal
propagating outwards from $r_{\rm e}$ to $r$ is obtained
using the stationary phase approximation
in complete analogy to the Minkowski case described
in detail in Sec.~V of Ref.~\cite{2020PhRvD.102d4010R}. The result of this
calculation is
\begin{eqnarray}
  \sigma(\eta,r) &=& A(\eta,r)
  e^{\iu \phi(\eta,r)}
  ~~~
  \text{with}
  \nonumber
  \\[10pt]
  \phi(\eta,r) &=&
  \sqrt{\Omega^2-\omega_*^2}(r-r_{\rm e})
  -\Omega \eta
  + \arg [\tilde{\sigma}(\Omega,r_{\rm e})]
  -\frac{\pi}{4}\,,
  \nonumber
  \\[10pt]
  A(\eta,r) &=&
  \sqrt{\frac{2}{\pi}}\,
  \frac{(\Omega^2-\omega_*^2)^{3/4}}
       {\omega_* \sqrt{r-r_{\rm e}}}
       |\tilde{\sigma}(\Omega,r_{\rm e})|\,,
  \nonumber
  \\[10pt]
  \Omega(\eta,r)
  &=&
  \frac{\omega_* \eta}{\sqrt{\eta^2-(r-r_{\rm e})^2}}\,,
  \nonumber
  \\[10pt]
  \omega_* &=& a\sqrt{\mu^2-H^2(1-q)}
  \approx a\mu\,.
  \label{eq:Aphom_etar}
\end{eqnarray}
Here, the last approximation arises from converting the
Hubble parameter into units where $\hbar=c=1$,
\begin{eqnarray}
 H\approx H_0 \approx 75~{\rm km\,s}^{-1}\,{\rm Mpc}^{-1} \approx 1.6\times 10^{-32}~{\rm eV} \nonumber
\end{eqnarray}
which is well below the scalar field's mass range $10^{-15}$ to $10^{-10}~{\rm eV}$
that we consider in this work.
Note also
that the observed signal $\sigma(\eta,r)$
in Eq.~(\ref{eq:Aphom_etar}) depends on the
emitted wave only through the presence of
$|\tilde{\sigma}(\Omega,r_{\rm e})|$ and
$\arg[\tilde{\sigma}(\Omega,r_{\rm e})]$ in
the amplitude $A$ and phase $\phi$. 

Our final task is to relate the conformal time $\eta$ and
the radial coordinate $r$ to the time and distance measurements
used by an observer on Earth, namely, the observer's proper
time $\tau$ and the source's luminosity distance $D_L$.
To this end, we first note that our coordinate
system $(\eta,r)$ is centered on the source with time chosen
such that
$\eta=0$ corresponds to the time of the collapse (which, we
recall, is treated as instantaneous on the cosmological timescale).
The luminosity distance between the source and observer
is given by the areal radius so that, at the time $\eta_{\rm o}$
of observation, $D_L=a(\eta_{\rm o}) r$.
Without loss of generality, we set the scale factor at the time of
emission to $a(\eta_{\rm e})=1$ so that
$a(\eta_{\rm o})=1+z$ with the standard redshift $z$.
Likewise, the observer measures the progression of her
age by $\du \tau =a(\eta_{\rm o})\du \eta$, with $\tau=0$
defined as the time of arrival of the {\it electromagnetic\/}
signal
from the supernova event. Taking into account that
$\eta=0$ marks the time of the collapse and bearing in
mind that\footnote{In practice, we extract
the source's GWs at $\mathcal{O}(1)$ light seconds from the
star's center which is negligible compared to the astrophysical
distances in the kpc or Mpc range.} $r_{\rm e} \ll r$, we find
\begin{equation}
  r=\frac{D_L}{a_{\rm o}}\,,
  ~~~~~
  \eta=\frac{\tau+D_L}{a_{\rm o}}\,.
\end{equation}
These are readily inserted into Eq.~(\ref{eq:Aphom_etar}) and yield
the GW signal of a core-collapse event at luminosity distance
$D_L$ at (proper) observer time $\tau$ elapsed since the
identification of the supernova event in electromagnetic
radiation,
\begin{eqnarray}
    \sigma(\tau,D_L) &=& A(\tau,D_L)e^{\iu \phi(\tau,D_L)}~~~~~\text{with}
  \nonumber \\[10pt]
  \phi(\tau,D_L) &=& -\mu\sqrt{(\tau+D_L)^2-D_L^2}
        +\arg [\tilde{\sigma}(\Omega,r_{\rm e})]
        - \frac{\pi}{4}\,,
  \label{eq:phitauDL}
  \nonumber \\[10pt]
  A(\tau,D_L) &=& \sqrt{\frac{2\mu}{\pi}}
  \frac{(1+z)D_L}{\left[
  (\tau+D_L)^2-D_L^2
  \right]^{3/4}}
  \left| \tilde{\sigma}(\Omega,r_{\rm e}) \right|\,,
  \label{eq:AtauDL}
  \nonumber \\[10pt]
  \Omega(\tau,D_L) &=& \frac{\omega_*}{\sqrt{1-\left(
        \frac{D_L}{\tau+D_L}\right)^2}}\,,
  \label{eq:OmegatauDL}
  \nonumber\\[10pt]
  \omega_* &=& (1+z)\mu \,.
\end{eqnarray}

Note that the last two equations should {\it not\/} be interpreted
as a blue shift of the signal. Rather, they demonstrate that
a mode reaching Earth at time $\tau$ requires a higher
frequency, i.e.,~larger group velocity, as compared to the
nonexpanding Minkowski case. In other words, the cosmological
expansion delays modes of lower frequency to a later arrival time,
just as we would intuitively expect.

\paragraph*{A note on massive amnesia~--~}
\label{sec:memory-effects}
Already in the early days of modeling GW emission from supernovae, a mechanism of GW production in the zero-frequency limit was discovered,
sourced by the outwards burst of relativistic neutrinos~\cite{Epstein:1978dv,Turner:1978jj}.
The presence of the resulting GW signal was confirmed by numerical simulations in recent decades, identified as a slowly increasing tail in the GW strain~\cite{Burrows:1995bb,Kotake:2006aq,Yakunin:2015wra}.
This feature can be interpreted as a linear ``memory effect,'' leaving a permanent relative displacement in the test masses of an interferometer.
Here, we observe a similar effect when extracting the waveform at short distances from the source (see Fig.~\ref{fig:dispersive_nature}), with the generation of the scalar field.
However, due to the massive dispersion relation, this feature, dominating the low-frequency end of the spectrum, does not propagate out to infinity and is therefore not measurable from a distant GW observatory.
Hence, no such observable feature is found in the scalar stochastic spectra studied here.

\section{Results}\label{sec:results}
\subsection{Core collapse}
\label{sec:corecollapse}

A systematic exploration of the space of parameters that
characterize the progenitor star, the EOS and the scalar field
has been performed
in Ref.~\cite{2020PhRvD.102d4010R}.
There, we observe a pattern in the dynamics and remnants obtained which leads to a classification of the core-collapse events into five different scenarios:
\begin{list}{\rm{(\arabic{count})}}{\usecounter{count}
             \labelwidth1cm \leftmargin1.0cm \labelsep0.4cm \rightmargin0cm
             \parsep0.5ex plus0.2ex minus0.1ex \itemsep0ex plus0.2ex}
\item Prompt collapse to a low-compactness, weakly scalarized NS which produces scalar radiation with an amplitude $\mathcal{O}(\alpha_0)$. We obtain this for less negative $\beta_0$ values for progenitors and EOSs which in the GR case result in the formation of NSs.
\item Prompt collapse to a high-compactness, strongly scalarized NS which produces scalar radiation
with an amplitude $\mathcal{O}(1)$. This scenario is
realized for sufficiently negative values of $\beta_0$,
but independent of the progenitor, EOS, and other scalar parameters.
\item Multistage collapse to a strongly scalarized NS which produces scalar radiation of magnitude
$\mathcal{O}(1)$. This occurs for an intermediate range of $\beta_0$ values slightly less negative
than those of case (2), independent of the progenitor model or EOS\@..
\item Two-stage collapse to a BH which produces scalar radiation of magnitude $\mathcal{O}(\alpha_0)$. We obtain this outcome for less negative $\beta_0$ values and progenitors and EOSs which, in the GR case, lead to BHs.
\item Multistage collapse (with at least three stages) to a BH which produces scalar radiation of amplitude $\mathcal{O}(1)$. This occurs for progenitors and EOSs which lead to BHs in the GR case and for an intermediate $\beta_0$ range between the values leading to a 2-stage BH formation and  those leading to multistage NS formation. 
\end{list}

\begin{figure}[t]
    \centering
    \includegraphics[width=0.48\textwidth]{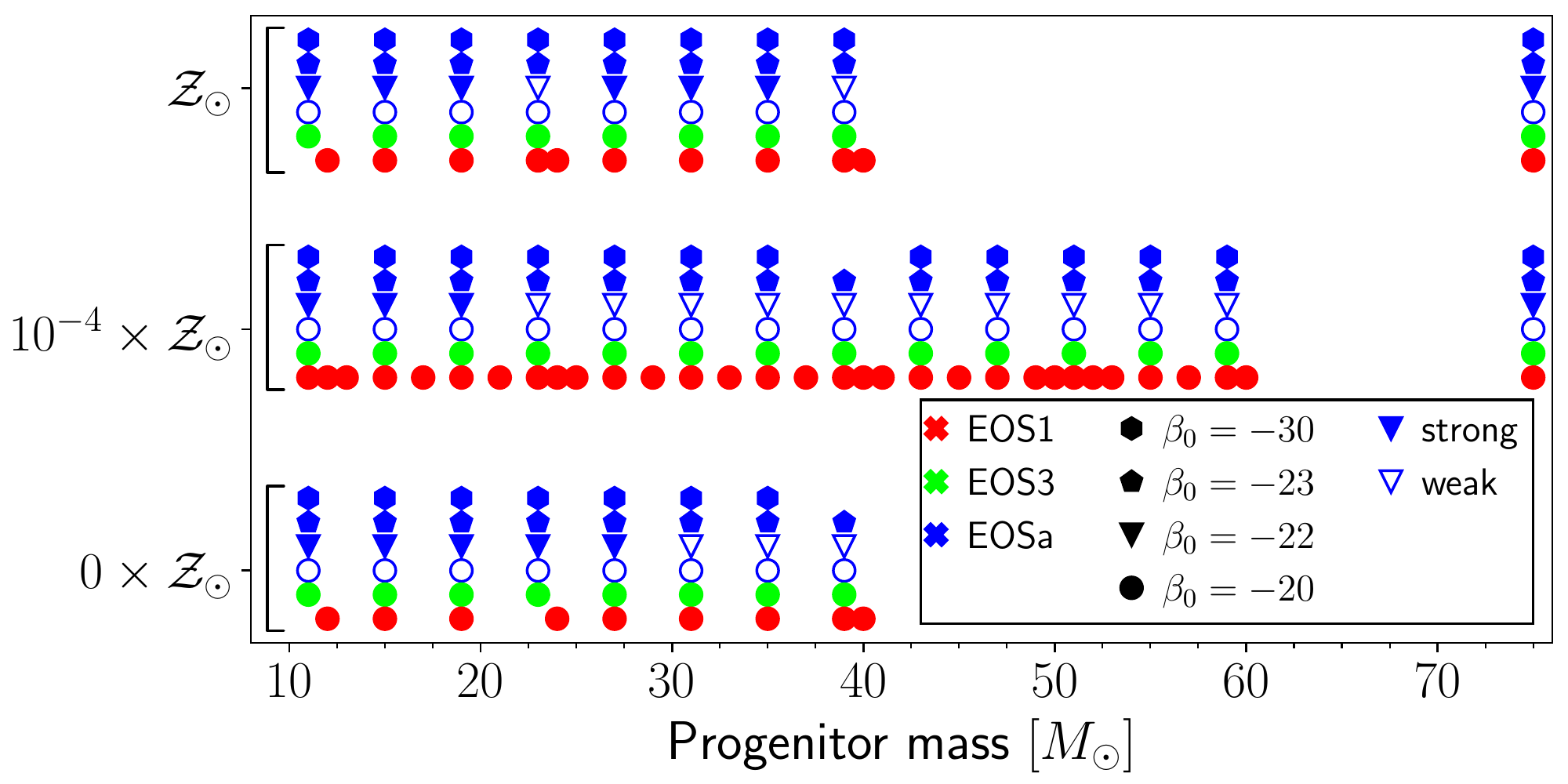}
    \caption{\label{fig:scalarization_level}
   Classification of scalarization based on the amplitude of the central scalar field for sets of simulations spanning a wide set of progenitors (filled markers for strong and empty markers for weak scalarization). The order of magnitude that these amplitudes reach in five of the six sets of simulations is plotted in Fig. 4 of~\cite{2020PhRvD.102d4010R}.
    }
\end{figure}

A more quantitative illustration of the $\beta_0$
intervals corresponding to these five main scenarios
is displayed in Fig.~3 of Ref.~\cite{2020PhRvD.102d4010R} for a progenitor star of $\rm M_{ZAMS}=39 \, M_\odot$, EOS3, and solar metallicity, 
and for a progenitor star of $\rm M_{ZAMS}=39 \, M_\odot$, EOS3, and primordial (zero) metallicity; in the
GR limit, these progenitor models collapse to
a NS and a BH, respectively. 

The main goal of this work is to assess how a large
number of core-collapse events of the type listed above
combines into generating a SGWB.
The different progenitors used as a starting point in our simulations are highlighted in Fig.~\ref{fig:scalarization_level}.
We mark the strength of the GW obtained for each configuration by using filled symbols for strong and empty symbols for weak radiation.
We use this finite grid of configurations as a basis to build our model for the stochastic background of scalar GWs.
As noted in Ref.~\cite{2020PhRvD.102d4010R}, there is no monotonic relation
between the progenitor mass and the threshold of strong scalarization.
This is best exemplified in the middle right panel of Fig.~4: for $\alpha=10^{-2},\,\beta_0=-22, \, \mu=10^{-14}$ eV, EOSa, and solar metallicity, several intermediate progenitor masses result in weak scalarization, whereas the rest are strongly scalarized.

\begin{figure*}
    \centering
    \includegraphics[width=0.99\textwidth]{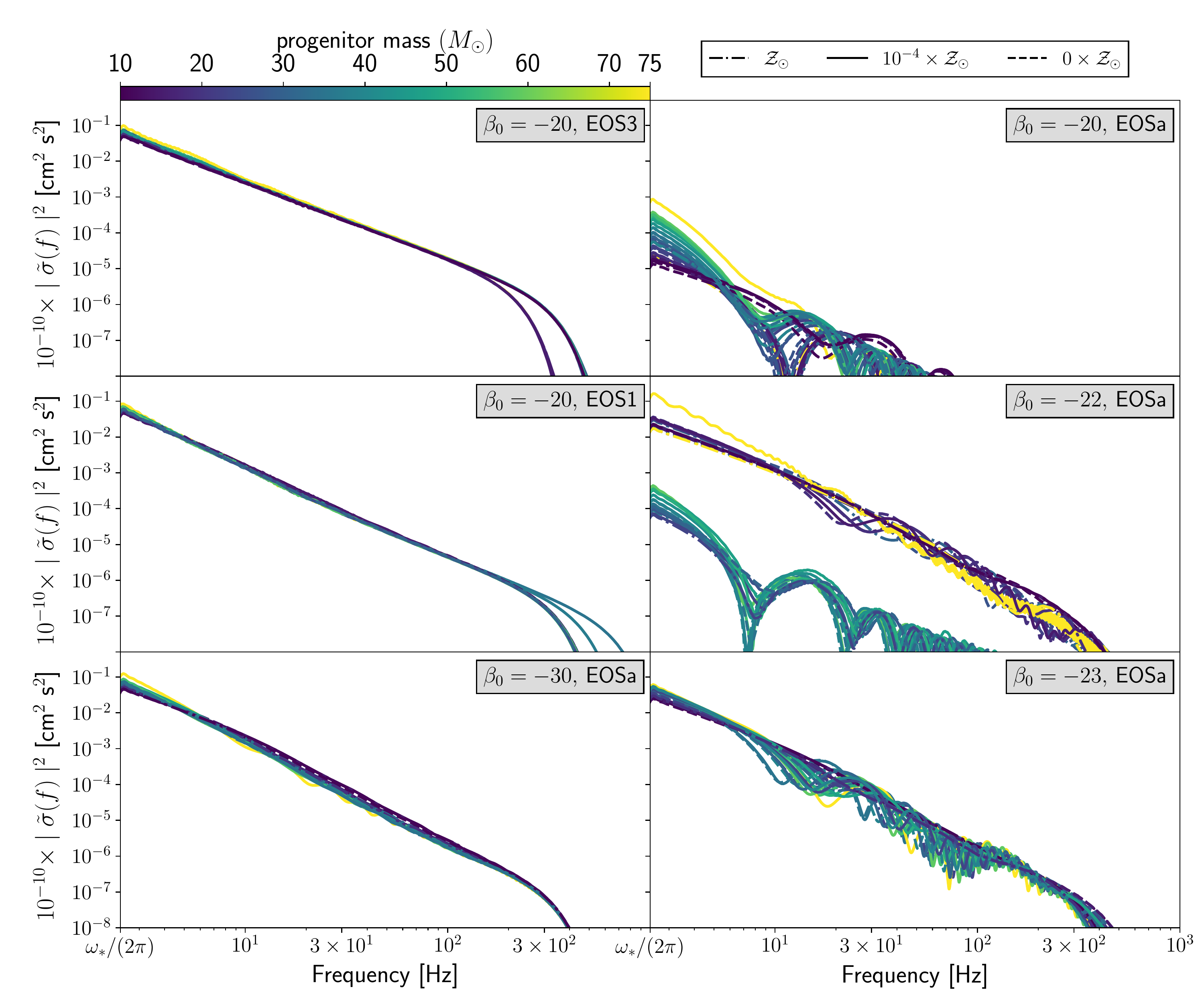}
    \caption{\label{fig:power_spectrums}
      Power spectra for each GW obtained using GR1D.
      We split the waveforms into six sets according to the EOS and $\beta_0$ values (the rest of the scalar parameters, $\mu=10^{-14}$ eV and $\alpha_0=10^{-2}$, are the same for all sets).
      The color of the line is parametrized by the mass of the progenitor as measured in $M_\odot$ (see color bar on top for reference).
      The metallicity is encapsulated in the line style as displayed in the top right legend. In the top two panels on the left, some differences appear for high frequencies, which are caused by the distinct resolution used in some of the simulations: higher resolution leads to minor improvements of the modeling of steep
gradients in the time domain, which results in slightly larger power at
high frequencies. Note, however, the logarithmic scale, so the
differences are small with no significant effect on the overall signal
power. The curves in the three left panels all correspond to scenario (2) of the
      list in Sec.~\ref{sec:corecollapse}, i.e.,~prompt collapse to strongly scalarized compact NSs, and all signals look essentially identical. On the right-hand side, all curves fall into two categories, low-amplitude signals for weakly scalarized remnants and
      high-amplitude signals (very much like those on the left) for
      strongly scalarized collapse remnants.
    }
\end{figure*}

The prompt high-compactness NS scenario of item (2) in the
above list is the simplest and, for our purposes,
most interesting case due to its universality: for any progenitor and EOS and a sufficiently negative $\beta_0$,
we always obtain a strongly scalarized neutron star and,
hence, a high-amplitude GW signal.
We have therefore selected, for our first computation of a stochastic background, scalar-field parameters that lead to this specific scenario: $\alpha_0=10^{-2},\, \beta_0=-20,\,\mu=10^{-14}$ eV for EOS1 and EOS3 and $\alpha_0=10^{-2},\, \beta_0=-30,\,\mu=10^{-14}$ eV for EOSa.
The universality of the resulting signals is demonstrated in Fig.~\ref{fig:power_spectrums} where we plot the resulting power spectra (see the left three panels of the figure). They show good agreement except for the low-amplitude
contributions at high frequency in the upper left
and center left panel, which are due to numerical
inaccuracy, and due to their small amplitude, they do not
contribute significantly to the overall power.

In the right panels of Fig.~\ref{fig:power_spectrums}, we highlight the behavior near the threshold of strong scalarization.
For this purpose we consider $\beta_0=-20,\,-22,\,-23$ for $\alpha=10^{-2}, \, \mu=10^{-14}$ eV, EOSa.
This is again best interpreted using the power spectrum of the GWs:
\begin{itemize}
    \item For $\beta_0=-20$, all progenitors result in the prompt formation of low-compactness NSs and GW signals 2 orders of magnitudes weaker than for $\beta_0 = -23$ (see top right panel of Fig.~\ref{fig:power_spectrums}).
  \item For $\beta_0=-22$, some simulations still produce weak scalar radiation (mostly intermediate mass progenitors), but several lead to strong scalarization (see center right panel of Fig.~\ref{fig:power_spectrums}).
        For this case, the power spectrum for the waveforms is less smooth as compared to the other scenarios displayed in the left panels.
        This behavior arises from the multistage character of the collapse to a NS and the correspondingly more complex structure of the GW signal in the time domain.
        A particularly strong signal is obtained for the progenitor model with mass $75\, M_\odot$ and $10^{-4}$ solar metallicity; the GW signal from this highly massive progenitor exceeds that of other strongly scalarized configurations by about a factor of 2.
    \item For $\beta_0=-23$, all simulations produce strong GW radiation (see bottom right panel of
    Fig.~\ref{fig:power_spectrums}), but again, some of their power spectra exhibit fluctuations due to the multistage NS scenario.
\end{itemize}

\subsection{Stochastic GW background}\label{sec:background}

GW signals from core-collapse supernovae (CCSN) in ST theory can be louder than in GR\@.
Therefore, they are detectable out to larger distances, increasing the rate of detectable events. 
Additionally, in \emph{massive} ST theories the dispersive nature of the wave propagation over astrophysically large distances stretches the signals out in time, perhaps by as much as several decades~\cite{2017PhRvL.119t1103S}.
This dramatically increases the signal duration and, coupled with the higher event rate, increases the probability that multiple signals will overlap in time.
If many such signals overlap, they can form a stochastic background of GWs.
In this section we present an order-of-magnitude estimate for the energy density in such a background, based on a representative subset of numerical simulations combined with realistic astrophysical population assumptions, and we consider the prospects for its detectability.

In ST theory the scalar field couples to the spacetime metric; see Eq.~\eqref{eq:action}.
Oscillations in the scalar field are detectable as GWs with a scalar polarization.
The strain amplitude of the GW signals is given by
\begin{equation}
h=2\alpha_0 \varphi\,,
\end{equation}
where $\alpha_0$ is the linear term in the coupling function; see Eq.~(\ref{eq:coupling}).
Here, we neglect the slight suppression in the GW strain amplitude at frequencies $\omega\gtrsim\omega_*$ from the presence of a longitudinally polarized GW in massive ST theory; see, e.g., Ref.~\cite{Rosca-Mead:2019seq}. 

The formalism for computing the expected energy density in a SGWB from a cosmological population of discrete sources has been studied by several authors; see, for example, Refs.~\cite{2001astro.ph..8028P, 2008MNRAS.390..192S}, and Ref.~\cite{2005PhRvD..72h4001B} in the specific context of supernovae.
The SGWB is commonly described in terms of its local energy density $\rho_{\rm GW}$ per logarithmic frequency interval, normalized to the critical cosmological density $\rho_c=(3H_0^2)/(8\pi)$: 
\begin{align}
    \Omega_{\rm GW} = \frac{1}{\rho_c}\frac{\ddd \rho_{\rm GW}(f)}{\ddd \ln f} .
\end{align}

\subsubsection{Astrophysical population statistics}
\label{sec:pop-stat}

Building a SGWB requires knowledge of the population of events across the Universe and how this population evolves over cosmic time.
We can parametrize this evolution by any cosmological distance parameter; here, we use the redshift $z$.
We estimate the rate $R(z)$ of CCSN events per unit comoving volume using three different models:
(i) the simple model described in Eq.~(7) of Ref.~\cite{2005PhRvD..72h4001B}, in which the rate increases, as we look back in time, as a power law in $(1+z)$ from its local value of $2 \times 10^{-4}$ Mpc$^{-3}$ yr$^{-1}$, followed by an abrupt transition to a constant for redshifts higher than $z=1$;
(ii) a rate that is directly proportional to the star-formation rate (SFR), as $R(z) = \lambda_{\rm CC} R_*(z)$, where the proportionality constant $\lambda_{\rm CC}\simeq 0.007 M_\odot$ is estimated based on the Salpeter initial mass function (IMF)~\cite{Salpeter:1955it} and the assumption that all stars above a threshold mass of $8M_\odot$ will eventually collapse~\cite{Finkel:2021zgf}. 
[the SFR function itself $R_*(z)$ is set to the model proposed by Springel and Hernquist, fit to observational data~\cite{Springel:2002ux,Hernquist:2002rg,Vangioni:2014axa}];
(iii) a rate similar to the latter, but with the SFR model changed to that of Madau and Dickinson~\cite{Madau:2014bja}.

\begin{figure}
    \centering
    \includegraphics[width=0.48\textwidth]{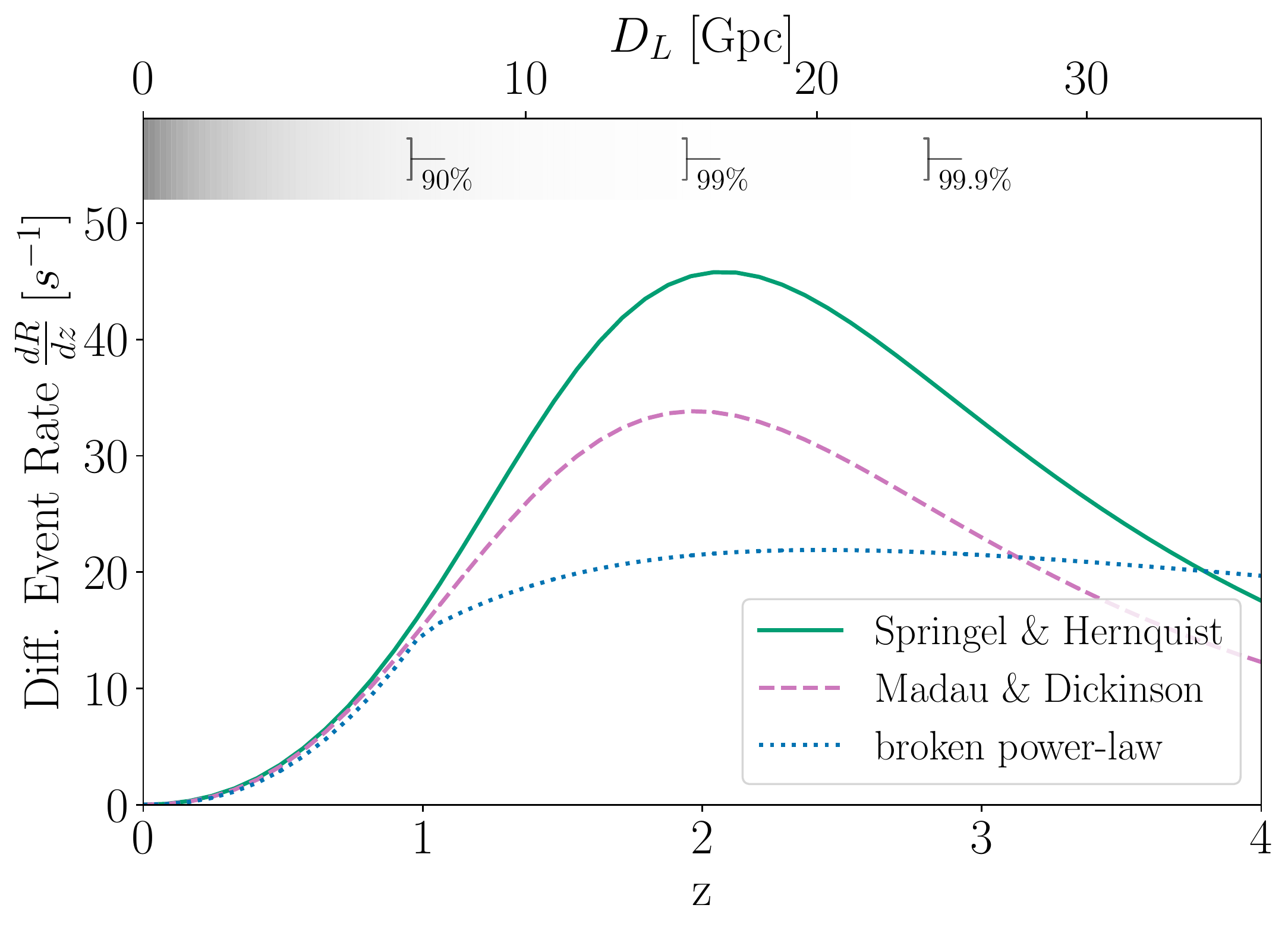}
    \caption{Three models for the core-collapse event rate per unit redshift, $\frac{\du R}{\du z} = R(z)\frac{\du V_{C}}{\du z}$, as a function of redshift.
      The intensity of the colored shading is proportional to the energy density contribution of the stochastic spectrum down to a given redshift at some fiducial frequency (here, $25$~Hz).
      The bracket markers indicate at what distances or redshifts the accumulated energy density (integrated from the observer outwards) crosses the given fractions of its total value.
    }
\label{fig:sfr}
\end{figure}

Since we only consider massive stars for our population (we set a universal threshold of $M_{\mathrm{ZAMS}}>8M_{\odot}$), the timescale of stellar evolution will be of the order of at most a few $\mathrm{Myr}$~\cite{Hurley:2000pk}, much smaller than the timescale of SFR variation.
Thus, any time lag between the SFR curve and the event-rate curve for core collapse of massive stars can be safely neglected.
The CCSN event rates predicted by the three models are depicted in the curves of Fig.~\ref{fig:sfr}, shaded by the (normalized) complementary-cumulative distribution (CCD) of $\Omega_{\rm GW}$ (integrated from $z$ to infinity). 
The shading and accompanying tick marks denoting the CCD values on the top of the plot are calculated based on the Springel and Hernquist curve (solid green) but are practically identical for all three models up to $z \sim 1$.
Evidently, the bulk of the contribution to the stochastic scalar spectrum comes from events at distances $z<1$.

\paragraph*{Astrophysical priors.---}
\label{sec:astro-priors}
To simulate a realistic population of stars, we sample the ZAMS mass from a power law following the Salpeter IMF, $p(m) \propto m^{-2.35}$, and the metallicity from a prior uniform in $\log_{10} Z \in [-6,-1]$.
We then map our randomly sampled stellar parameters onto the nonuniform discrete set of points corresponding to our numerical simulations, by means of a Voronoi tessallation of the $M_{\mathrm{ZAMS}}-\log Z$ plane using the Manhattan distance, $d(u^i,v^j) = \sum_{i} |u^{i} - v^{i}|$.

The prior just described is separable; meaning that it is not able to accommodate any correlation between ZAMS mass and metallicity. Astrophysically, there appears to be little correlation between stellar metallicity at different star-forming environments and the IMF~\cite{Myers:2011at}.
We should, however, expect some level of correlation between ZAMS mass and metallicity, and we should also expect that this relation itself will vary across cosmic history as stars are being formed in an increasingly metal-rich environment.
The accuracy of the prior and the effect of neglecting this correlation have been checked by repeating the analysis using the alternative mass-metallicity relationship of~\cite{Ma:2015ota} or~\cite{Langer:2005hu}; we find that the effect of this redshift dependence on the final scalar stochastic spectrum is minimal, when compared to the uncertainties due to the sparsity of our simulations on the parameter space.

\subsubsection{Energy spectrum from CCSN events}
\label{sec:gw-energy-spectrum}

Building a SGWB also requires knowledge of the GW signals produced by each supernovae source, as a function of frequency in the observer's frame.

The energy density in the GW background at any given frequency $f$ is given by the integral of all plane-wave contributions from distance sources of GW radiation at that frequency, in the observer's frame.
The contribution from each type of event (of given mass and metallicity) comes from its energy spectrum emitted in GW radiation,
\begin{eqnarray}
\label{eq:dEdt}
\frac{\mathrm{d} E_{\mathrm{GW}}}{\mathrm{d} f_s} & = & \frac{c^3 (2\pi f_s)^2}{16\pi G} \int \left\langle (\tilde{h}_+)^{2} + (\tilde{h}_\times)^{2} + (\tilde{h}_S)^{2} \right\rangle \mathrm{d}\Omega \nonumber \\
& = &  \frac{c^3\pi^2f_s^2}{G} \left\langle  {\tilde{h}_{S}(f_s)}^{2} \right\rangle\,,
\end{eqnarray}
where in the last step we have assumed spherical symmetry.

Within a comoving volume $\mathrm{d}V_C(z)$ enclosed within a shell between cosmological redshifts $[z,z+\mathrm{d}z]$, we have $\mathrm{d}V_{C}(z) R(z)$ events of intrinsic radiated energy density $\mathrm{d} E_{\mathrm{GW}}/\mathrm{d} f_{s}$, whose aggregated signal is observed at a luminosity distance $D_L(z)$ and redshifted as $f = f_s (1+z)$.

For a population of supernovae evolving throughout cosmic time, the GW spectrum is then given by
\begin{align} \label{eq:cosmo_back_GW}
  \Omega_{\rm GW}(f) =
  \int \frac{\ddd z}{1+z}
  \int \ddd \mathbf{\theta}
  \frac{4\pi}{15 \rho_c}\frac{\mathrm{d} R(z)}{\mathrm{d} \mathbf{\theta}} \frac{\ddd V_C}{\ddd z} \frac{\ddd t}{\ddd z} f_s^3 |\tilde{h}_S(f_s; \mathbf{\theta})|^2 ,
\end{align}
where $\frac{\mathrm{d} R(z)}{\mathrm{d} \mathbf{\theta}}$ is the rate of supernova events per comoving volume as a density in the mass-metallicity plane and $\tilde{h}_S(f; \mathbb{\theta})$ is the scalar GW signal of an individual event,
both estimated for some stellar parameters $\mathbf{\theta} = \{M_{\rm ZAMS},Z\}$ and at a redshift $z$.
The effects of cosmology enter through the observed event rate via $\ddd V_C(z)$ and via the factor $\ddd t/\ddd z$.
For a spatially flat FLRW geometry (with negligible radiation content) we have
\begin{align}
    \frac{\ddd V_c}{\ddd z} & = 4 \pi \frac{1}{H(z)} \left(\int_{0}^{z} \frac{1}{H(z')} \ddd z' \right)^2 \, , \,
    \frac{\ddd t}{\ddd z} = \frac{1}{(1+z)H(z)} \, , \nonumber \\
    H(z) & = H_0 \sqrt{\Omega_m (1+z)^3+\Omega_\Lambda}.
\end{align}
For the cosmological parameters $H_0$, $\Omega_m$ and $\Omega_\Lambda$, we use the Planck 2018 values~\cite{Planck:2018vyg}.

We first consider our baseline model case where the parameters of the massive ST theory are $\mu=10^{-14}~{\rm eV}$,
$\alpha_0=10^{-2}$, and $\beta_0=-20$, and the equation of state is EOS1; to the best of our knowledge, this theory is compatible with existing astrophysical constraints.
In this case, the GW signal from the supernova does not depend sensitively on the astrophysical parameters (e.g., stellar mass and metallicity) of the progenitor star; this can be seen in the center left panel of Fig.~\ref{fig:power_spectrums}.
Therefore, we can safely assume that in their rest frames every supernova produces an almost identical GW signal, while the superposition from the population can be estimated by taking the average of the curves in the center left panel of Fig.~\ref{fig:power_spectrums}.

\begin{figure}
    \centering
    \includegraphics[width=0.48\textwidth]{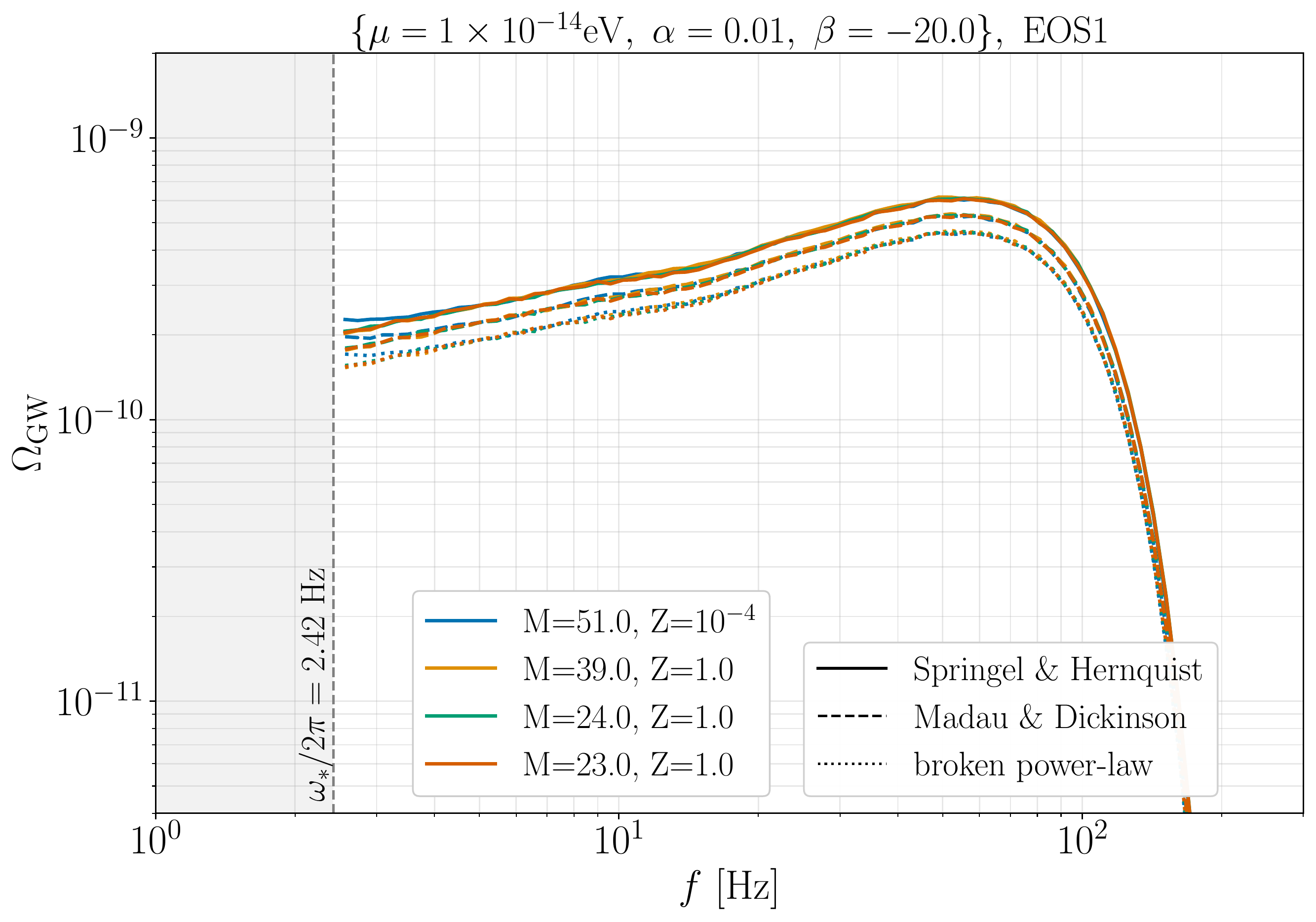}
    \caption{Energy density spectrum for the SGWB of the massive scalar field, produced by an astrophysical population of stellar core-collapse events. The theory parameters and EOS for our baseline model shown here are given at the top of the figure.
    For each of the four configurations considered (see color legend), we estimate the spectrum for the three different models of the event rate.
    Our model exhibits a sharp feature at $\omega_*$, the low-frequency cutoff of our massive scalar spectrum.
    }
\label{fig:omggw}
\end{figure}

After evaluating the integral of Eq.~\eqref{eq:cosmo_back_GW}, where, for each of the four different configurations we keep the model $\tilde{h}(f)$ constant across masses and metallicities (i.e.\ remove the $\mathbf{\theta}$ dependence), we arrive at predictions for the energy density in the scalar-polarized GW background in our baseline theory, which we plot in Fig.~\ref{fig:omggw}.
As expected, the stochastic GW spectra are practically identical, and the predominant source of uncertainty comes from modeling the event rate.
We can therefore conclude that for our baseline model, the CCSN-induced scalar SGWB peaks at $\sim 60$ Hz, with a peak energy density of
\begin{align}
\label{eq:loudest_baseline}
    \Omega_{\rm GW}(f=60\,\mathrm{Hz}) \approx 6\times 10^{-10}.
\end{align}

\begin{figure*}
    \centering
    \includegraphics[width=0.98\textwidth]{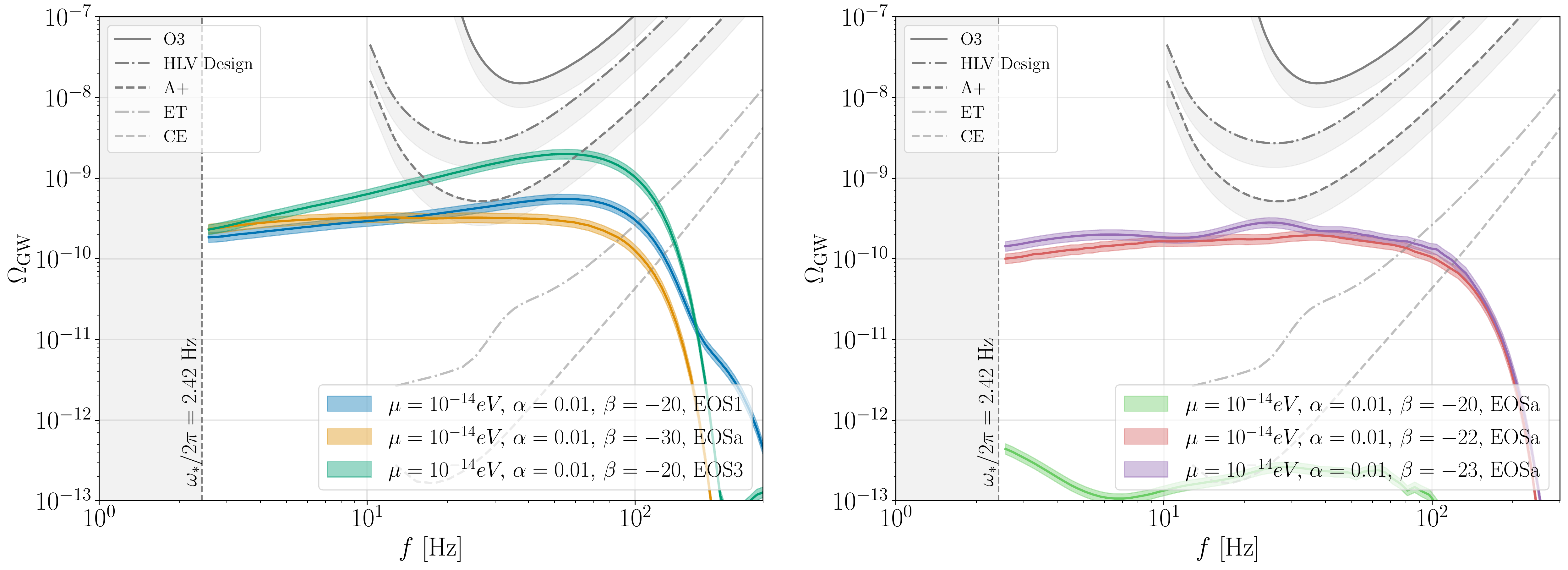}
    \caption{Energy density spectra of the scalar stochastic background from stellar core collapse, for a selection of massive scalar-tensor theory parameters and EOSs.
    The sensitivity curves for HLV in O3, and one year of HLV at design sensitivity, A+, ET, and CE are shown in gray; the gray shading extends the sensitivity to about two years of operation or, equivalently, a $1\sigma$ detection with one year of data (see text for details).
    The error bands of the stochastic spectra represent the uncertainty in modeling the astrophysical event rates for CCSNe.
    }
\label{fig:omg_gw_theories}
\end{figure*}

The above universality does not hold for all choices of theory parameters and EOSs, as can be clearly seen in the right panels of Fig.~\ref{fig:power_spectrums}, especially for the most interesting cases where for some values of mass and metallicity, the star undergoes strong scalarization while for others it does not.
In Fig.~\ref{fig:omg_gw_theories}, we show the scalar SGWB spectra for each of the six ST models considered here, after weighting the entire set of simulated configurations with the appropriate astrophysical priors and performing the full parameter-dependent integral of Eq.~\eqref{eq:cosmo_back_GW}.
The left and right panels correspond to the models in the left and right columns of Fig.~\ref{fig:power_spectrums}, respectively.
We observe that, above the sharp low-frequency cutoff at $\omega_{*}$, the stochastic spectrum tends to rise slowly until $\sim100$ Hz and rapidly falls off beyond that range.
An exception with more erratic behavior is the case of $(\beta=-20, \mathrm{EOSa})$, with a much weaker spectrum.
This is to be expected since none of the simulated configurations in that theory shows signs of strong scalarization (see Fig.~\ref{fig:scalarization_level}).
Among the theories that produce strongly scalarized cores, we find that $(\beta=-20,~\mathrm{EOS3})$ gives the strongest stochastic signal, peaking at $\sim60$ Hz, with a peak energy density of
\begin{align}
\label{eq:loudest}
\Omega_{\mathrm{GW}}(f=60\, \mathrm{Hz}) \approx 2 \times 10^{-9},
\end{align}
and a slightly lower value at $25$ Hz, where the (LIGO) Hanford, (LIGO) Livingston, Virgo (HLV) network reaches its best sensitivity.
This value is smaller than current constraints on a Gaussian SGWB (see, e.g., Refs.~\cite{PhysRevLett.120.201102, 2019PhRvD.100f1101A}, \cite{KAGRA:2021kbb}) but likely detectable with two years of the LIGO-Virgo-KAGRA network at design sensitivity~\cite{KAGRA:2013rdx}.

In Fig.~\ref{fig:omg_gw_theories}, the sensitivity curves for a network of ground-based interferometric detectors are plotted, for the detection of a tensorial power-law spectrum at the 95\% credible level with one year of collected data.
The gray regions extend to the sensitivity of approximately two years of operation of the same network, again at the 95\% credible level (or, equivalently, a $1\sigma$ detection with one year of data).
The corresponding sensitivities for a scalar signal may be reduced by a factor of a few, depending on the frequency.
The empirical value range for this factor with the current network of detectors can be seen by comparing the bounds obtained for scalar-polarized and tensor-polarized SGWBs with O1, O2, and O3 data~\cite{Thrane:2013oya,PhysRevLett.120.201102,2019PhRvD.100f1101A,KAGRA:2021kbb}.
For the theories studied here, the expected scalar SGWB will be well probed by the planned next generation of ground-based detectors, namely the Einstein Telescope~\cite{Maggiore:2019uih} and the Cosmic Explorer~\cite{Evans:2021gyd},
down to $O(10^{-12})$ in the range of tens of Hz, or even with a few years of data from LIGO $A+$ at design sensitivity~\cite{2016PhRvL.116m1102A,KAGRA:2013rdx,Aplus:2018}.
A nondetection at that sensitivity would be strong evidence against the massive ST modified theory of gravity for a range of parameters that are compatible with all existing astrophysical constraints.
This result implies that it might be possible within the next decade to place new constraints on the parameter space of massive ST gravity using upper limits on the stochastic background of GWs with scalar polarizations.

We should note that the curves for the next-generation network of detectors are shown as extrapolations, based on the ones described above for the HLV network, on which we perform a rescaling equal to the PSD ratio at each frequency.
These are only indicative order-of-magnitude projections and not accurate estimates, as the actual power-law sensitivity curve is estimated as the envelope of the family of upper bounds for each value of the exponent~\cite{Thrane:2013oya} (straight lines in a logarithmic plot), and therefore, the resulting power-law sensitivity curve should always be convex.
Furthermore, the true power-law sensitivity estimate requires exact knowledge of the network geometry, so the overlap-reduction function (ORF) between detectors in the network can be calculated (see~\cite{Allen:1997ad,Maggiore:2007ulw} and~\cite{Babusci:2001tw,Nishizawa:2009bf} for the nontensorial calculations).
For instance, the triangular collocated interferometers in the ET observatory (here we use the ET-D noise curve~\cite{Hild:2010id}) will probe correlations between their data down to much shorter wavelengths.
Thus, unlike the HLV network for which the ORFs are quickly damped to zero at frequencies above $\sim100$ Hz, the ET configuration will be described by an ORF that is practically flat across the entire sensitivity band.

Some care should also be taken when comparing our predicted energy density to these observational constraints. These constraints are generally derived assuming the background satisfies Gaussian statistics.
For a background formed from a finite number of discrete sources, this will not be a good description.
We have only calculated the expected energy density in the background; we have not considered the extent to which it is Gaussian.

\section{Discussion and Conclusions}\label{sec:discussion}

The core-collapse and spontaneous scalarization processes in ST gravity are efficient sources of scalar GWs. 
If the scalar is massless, then these are burst sources of GWs.
However, for massive scalars, the stretched signals are continuous sources of GWs. 
In some regions of the ST theories' parameter space where hyperscalarization occurs, signals can be detected out to cosmological distances where the supernova rate is high enough to give many overlapping sources, thereby forming a stochastic GW signal.
Therefore, ST theories can be constrained observationally using results from the burst, continuous wave, and stochastic GW analyses now being performed by the LIGO and Virgo collaborations.

In this work, we focus on the SGWB arising from supernovae. To this end, we have performed a large number of 1+1 dimensional core-collapse simulations, covering a range of astrophysical parameters and EOSs. For a large part of the parameter space of these theories, the GWs display a universal behavior.

With realistic estimates of the event rate and NR-based models for the energy density spectrum, we have modeled the stochastic background of scalar GWs from CCSN events out to cosmological distances. To this end, we have generalized the modeling of the
propagation of massive signals in the framework of the
stationary phase approximation
to the case of $k=0$ FLRW cosmological spacetimes.

The estimated values within the LIGO or Virgo sensitivity range are smaller than current constraints; they will be measurable when the detectors operate at design sensitivity and for future generation detectors (ALIGO+, Einstein Telescope, Cosmic Explorer), allowing us to probe the theory's parameter space even deeper.

Regardless of whether the scalar SGWB described in this work is present or not, searches for a stochastic signal may well return positive results in the not-too-distant future.
A prime example of such a plausible scenario is the detection of the stochastic background coming from the population of inspiraling compact binaries, whose frequency and estimated amplitude largely overlap with the strong scalarization scenarios studied here (see, e.g., Fig.~23 of
Ref.~\cite{LIGOScientific:2021psn} for predictions using the latest binary population estimates).
In fact, searches with pulsar timing array (PTA) data in the nHz range are already showing the first weak evidence for the presence of a stochastic signal with gradually increasing confidence~\cite{NANOGrav:2020bcs,Goncharov:2021oub}.
Interestingly enough, there is little evidence for compatibility with the Hellings-Downs curve that defines the angular correlation of a tensorial transverse signal; a scalar transverse signal seems to be preferred by an odds ratio of at least 20:1, if a stochastic signal is at all present~\cite{NANOGrav:2021ini,Chen:2021wdo,Chen:2021ncc,Wu:2021kmd}.

The stochastic GW signal presented here, due to scalarization of newly formed neutron stars in CCSN events, can, in principle, be distinguished from other types of stochastic backgrounds of astrophysical or cosmological origin.
This can be achieved by examining (i) its pure scalar polarization content (GR does not permit GWs with scalar polarization); (ii) its angular distribution across the sky, which should follow the stellar density distribution at low redshifts; and (iii) the presence of a sharp spectral feature in the form of the characteristic low-frequency cutoff.
The frequency of this cutoff gives a direct measurement of the mass of the scalar field.


\begin{acknowledgments}
We thank Isobel Romero-Shaw and Lieke van~Son for insightful discussions on astrophysical population models.
R.R.M. acknowledges support by the Deutsche Forschungsgemeinschaft (DFG) under Grants No. 2176/7-1 and No. 406116891 within the Research Training
Group RTG 2522/1.
M.A. is supported by the Kavli Foundation.
This work has been supported by
STFC Research Grant No. ST/V005669/1,
``Probing Fundamental Physics with Gravitational-Wave Observations,''
and
NSF Grant No.~PHY-090003. This research project was conducted using
computational resources at the
Maryland Advanced Research Computing Center
(MARCC)
as well as the Cambridge Service for Data Driven Discovery (CSD3)
system at the University of Cambridge
and Cosma7 and 8 of Durham University
inside the DiRAC allocation ACTP284
through STFC capital Grants
No.~ST/P002307/1 and No.~ST/R002452/1, and STFC operations Grant No.~ST/R00689X/1.
We made use of presupernova models by S. Woosley and A. Heger available at \href{https://2sn.org/stellarevolution/}
{2sn.org/stellarevolution}.
This work made use of the following publicly available Python packages: \texttt{astropy} (an ecosystem of tools and resources for astronomy \cite{Astropy:2022ucr}), \texttt{numpy}~\cite{Harris:2020xlr}, \texttt{scipy}~\cite{Virtanen:2019joe}, and \texttt{matplotlib}~\cite{Hunter:2007ouj}.
\end{acknowledgments}


\bibliographystyle{apsrev4-2}
\bibliography{bibliography}


\appendix
\section{Details of the wave propagation}\label{app:A}
In this appendix, we derive in more detail the
expressions (\ref{eq:Aphom_etar}) for the gravitational wave signal after
propagation across large distances in our cosmological
neighborhood. Our starting point for this calculation
is the Klein-Gordon equation (\ref{eq:wavetime}) in the
time domain for a scalar field on a spatially flat
Friedmann-Lema{\^\i}tre-Robertson-Walker (FLRW) spacetime given
by the line element (\ref{eq:ds2_FLRW}). Fourier transforming the
wave equation according to Eq.~(\ref{eq:FT}) gives
\begin{eqnarray}
  &&\omega^2 \tilde{\sigma}(\omega,r)
  + \int_{-\infty}^{\infty}
  a^2H^2(1-q)\sigma
  e^{\iu \omega \eta} \du \eta
  +
  \partial_r^2 \tilde{\sigma}(\omega,r)
  \nonumber \\[10pt]
  &&
  \hphantom{
  \omega^2 \tilde{\sigma}(\omega,r)
  }
  =
  \int_{-\infty}^{\infty}
  \mu^2 a^2 \sigma e^{\iu \omega \eta}\du \eta\,,
  \label{eq:waveFTaux}
\end{eqnarray}
where $a$, $H$ and $q$ are functions of conformal time
$\eta$.

We now assume that at any given radius, the duration of the signal
$\sigma$ is short compared to the timescale of cosmological
evolution. In that case, the scale factor $a(\eta)$ and its
derivatives do not vary significantly during the passage of
the signal, and we can approximate $a$, $H$, and $q$ as
constants in the integrals of Eq.~(\ref{eq:waveFTaux}). This
leads to the simplified wave equation (\ref{eq:waveFT1})
with the general solution (\ref{eq:waveFTgensol}),
which we repeat here for convenience,
\begin{widetext}
\begin{eqnarray}
  &&
  \tilde{\sigma}(\omega,r) = \tilde{f}(\omega)
  e^{\iu k(r-r_{\rm e})}
  +
  \tilde{g}(\omega)
  e^{-\iu k(r-r_{\rm e})}
  ~~~~~\text{with}~~~~~
  k=\sqrt{\omega^2-\omega_*^2}
  \nonumber \\[10pt]
  &\Rightarrow&
  \sigma(\eta,r)
  =
  \frac{1}{2\pi}
  \int_{-\infty}^{\infty}
  \tilde{f}(\omega)
  e^{\iu [k(r-r_{\rm e})-\omega \eta]}
  +
  \tilde{g}(\omega)
  e^{-\iu[k(r-r_{\rm e})+\omega\eta]}
  \,\du \omega\,.
  \nonumber
\end{eqnarray}
Introducing the radial variable $\varrho \defeq r-r_{\rm e}$,
this solution consists of the Fourier modes
\begin{equation}
  \tilde{f}(\omega)e^{\iu (k\varrho-\omega \eta)}\,,~~~~~
  \tilde{g}(\omega)e^{-\iu(k\varrho + \omega \eta)}\,.
\end{equation}
We now have three regimes for the frequency $\omega$,
\begin{eqnarray}
  \omega>\omega_*:&&~~~ k \in \mathbb{R}\,,~~~
  \tilde{f}\text{ outgoing},~~~
  \tilde{g}\text{ ingoing}
  \,, \nonumber \\[10pt]
  -\omega_* < \omega < \omega_*:&&~~~
  \iu k \in \mathbb{R}\,,~~~
  \tilde{f}(\omega)e^{\iu(k\varrho-\omega \eta)}
  \propto e^{-|k|\varrho}\,,~~~
  \tilde{g}(\omega)e^{-\iu (k\varrho+\omega\eta)}
  \propto e^{|k|\varrho}
  \,,\nonumber \\[10pt]
  \omega<-\omega_*:&&~~~
  k\in\mathbb{R}\,,~~~
  \tilde{f}\text{ ingoing},~~~
  \tilde{g}\text{ outgoing}\,.
\end{eqnarray}
Imposing the requirement that our signal is real and bounded,
we find
\begin{eqnarray}
  &&\text{for}~~|\omega| > \omega_*:~~~
  \tilde{g}^*(-\omega) = \tilde{f}(\omega)\,,
  \nonumber \\[10pt]
  &&\text{for}~~|\omega|<\omega_*:~~~
  \tilde{g}(\omega)=0~~~\text{and}~~~
  \tilde{f}^*(-\omega)=\tilde{f}(\omega)\,.
  \label{eq:conditionsreal}
\end{eqnarray}
Finally, we assume that the signal does not contain a standing
wave, so $\tilde{f}(\pm \omega_*)=\tilde{g}(\pm\omega_*)=0$.
With these conditions, the time domain solution becomes
\begin{equation}
  2\pi \sigma(\eta,r)
  =
  \int_{\Sigma}\tilde{f}(\omega)e^{\iu(k\varrho-\omega\eta)}
  +\tilde{g}(\omega)e^{-\iu(k\varrho+\omega \eta)}\,\du \omega
  +
  \int_{\bar{\Sigma}}
  \tilde{f}(\omega)
  e^{-|k|\varrho}\,e^{-\iu \omega\eta}\,\du \omega\,,
  \label{eq:waveIFT}
\end{equation}
with the intervals $\Sigma\defeq (-\infty,-\omega_*)\cup
(\omega_*,\infty)$ 
and $\bar{\Sigma}\defeq (-\omega_*,\omega_*)$.
Because of its exponential decay with $\varrho$, the second integral
becomes negligible at astrophysical distances, leaving us with
the integral over $\Sigma$. For its evaluation, we first
restore the integration domain to $\mathbb{R}$ by formally
setting $\tilde{f}(\omega)=\tilde{g}(\omega)=0$ over $(-\omega_*,
\omega_*)$. Second, we apply the {\it stationary-phase approximation\/}
(SPA) whereby for small $\epsilon$, integrals of the form
\begin{equation}
  I(\epsilon)\defeq \int_{-\infty}^{\infty} A(\omega)
  e^{\iu \vartheta(\omega)/\epsilon}\,\du \omega
  \label{eq:SPA}
\end{equation}
are dominated by frequencies $\Omega$ where $\vartheta'=0$.
This allows us to Taylor expand
\begin{eqnarray}
  && \vartheta(\omega) = \vartheta(\Omega)
  +\underbrace{\vartheta'(\Omega)}_{=0}(\omega-\Omega)
  +\frac{1}{2} \vartheta''(\Omega)(\omega-\Omega)^2 + \ldots\,,
  \nonumber \\[10pt]
  && A(\omega) = A(\Omega) + \ldots\,,
\end{eqnarray}
so that, using the substitution $s=\omega-\Omega$,
\begin{equation}
  I(\epsilon)
  \approx
  A(\Omega)
  e^{\iu \vartheta(\Omega)/\epsilon}
  \int_{-\infty}^{\infty}
  \exp
  \left[
  \frac{\iu \vartheta''(\Omega)}{2\epsilon}s^2
  \right]
  \du s
  \,.
  \label{eq:SPA2}
\end{equation}
If our
function $\vartheta(\omega)$ has more than one extremum, we add up
the individual contributions. We next write the time-domain wave
signal (\ref{eq:waveIFT}) in the form
\begin{equation}
  2\pi \sigma(t,r) = \int_{-\infty}^{\infty}
  \left[
  \tilde{f}(\omega)e^{\iu \vartheta(\omega)\varrho}
  +\tilde{g}(\omega)e^{\iu\theta(\omega)\varrho}
  \right]\,\du \omega\,,~~~
  \text{with}~~~
  \vartheta(\omega) = k-\omega \frac{\eta}{\varrho}\,,~~
  \theta(\omega)=-k-\omega \frac{\eta}{\varrho}\,,~~
  k=\sqrt{\omega^2-\omega_*^2}\,.
  \label{eq:sigma3}
\end{equation}
Comparing with Eq.~(\ref{eq:SPA}), we identify $\epsilon=1/\varrho$
and, introducing the velocity $v\defeq \varrho/\eta$, find
\begin{eqnarray}
  &&\vartheta'(\omega)=0
  ~~~\Rightarrow~~~
  v = \left(
  \frac{\du k}{\du \omega}
  \right)^{-1}
  ~~~\Rightarrow~~~
  \omega = \mathrm{sign}(v)\frac{\omega_*}{\sqrt{1-v^2}}
  = \mathrm{sign}(v)\Omega
  ~~~\text{for}~~~
  \Omega = \frac{\omega_*}{\sqrt{1-v^2}}\,,
  \nonumber \\[10pt]
  &&\theta'(\omega)=0
  ~~~\Rightarrow~~~
  v = -\left(
  \frac{\du k}{\du \omega}
  \right)^{-1}
  ~~~\Rightarrow~~~
  \omega = -\mathrm{sign}(v) \Omega\,.
\end{eqnarray}
Of course, $v$ is the group velocity $\pm\du \omega/\du k$,
which we thus identify as a direct consequence of the SPA\@.
Note that for positive $v$, we have a contribution through
$\vartheta(\omega)$ at positive frequency $\Omega$ and
through $\theta(\omega)$ at negative frequency $-\Omega$.
The reverse is true for negative velocity, in agreement with our
earlier interpretation of the in- or outgoing nature of $\tilde{f}$ and $\tilde{g}$.

From now on, we restrict ourselves to positive $v$,
i.e., outgoing radiation, which
is the scenario relevant for the GW emission from core-collapse
events. The corresponding treatment of ingoing radiation
proceeds in complete analogy, differing only in some
sign flips due to the negative velocity. For $v>0$, we distinguish
two cases. For $v>1$, there exist no extrema of $\vartheta(\omega)$
and $\theta(\omega)$, so $\sigma$ vanishes;
unsurprisingly, there is
no superluminal radiation. For $v<1$, we have two extrema,
$\vartheta'(\Omega)=0$ and $\theta'(-\Omega)=0$, so Eq.~(\ref{eq:sigma3}) with the SPA (\ref{eq:SPA2}) becomes
\begin{equation}
  2\pi \sigma(\eta,r) = \tilde{f}(\Omega)
  e^{\iu \vartheta(\Omega)\varrho} I_f
  + \tilde{g}(-\Omega)
  e^{\iu \theta(-\Omega)\varrho} I_g
  \label{eq:SPA1}
\end{equation}
with
\begin{eqnarray}
  I_f = \int_{-\infty}^{\infty}
  e^{\frac{\iu(\omega-\Omega)^2}{\epsilon_f}} \du \omega
  = \int_{-\infty}^{\infty}
  e^{\frac{\iu\omega^2}{\epsilon_f}} \du \omega
  \,,~~~
  \epsilon_f = \frac{2}{\vartheta''(\Omega)\varrho}
  \,,\nonumber \\[10pt]
  I_g = \int_{-\infty}^{\infty}
  e^{\frac{\iu(\omega+\Omega)^2}{\epsilon_g}} \du \omega
  = \int_{-\infty}^{\infty}
  e^{\frac{\iu\omega^2}{\epsilon_g}} \du \omega
  \,,~~~
  \epsilon_g = \frac{2}{\theta''(-\Omega)\varrho}\,.
\end{eqnarray}
Using
\begin{equation}
  \vartheta''(\Omega) = \frac{-\omega_*^2}{k_0^3}\,,
  ~~~~~
  \theta''(-\Omega) = \frac{\omega_*^2}{k_0^3}\,,
  ~~~~~
  k_0\defeq k(\Omega)=\sqrt{\Omega^2-\omega_*^2}\,,
\end{equation}
we find $\epsilon_f=-\epsilon_g<0$ and, with the Fresnel
integral
\begin{equation}
  \int_{-\infty}^{\infty} e^{\iu t^2/\varepsilon} \du t
  =
  \sqrt{\frac{\pi | \varepsilon|}{2}}
  [1+ \mathrm{sign}(\varepsilon)]\,,
\end{equation}
we obtain
\begin{equation}
  I_f =
  \sqrt{\frac{\pi k_0^3}{\varrho \omega_*^2}}(1-\iu)\,,
  ~~~~~
  I_g =
  \sqrt{\frac{\pi k_0^3}{\varrho \omega_*^2}}(1+\iu)\,.
\end{equation}
Combining this result with Eq.~(\ref{eq:SPA1}), using
$\theta(-\Omega)=-\vartheta(\Omega)$, and
recalling the conditions (\ref{eq:conditionsreal}),
the time domain signal at the observer becomes
\begin{equation}
  2\pi \sigma(\eta,r)=
  2\myRe\left\{
  \tilde{f}(\Omega)e^{\iu \vartheta(\Omega)\varrho}
  \sqrt{\frac{\pi k_0^3}{\varrho \omega_*^2}}
  (1-\iu)
  \right\}\,.
\end{equation}
Note that the signal is now given exclusively in terms
of positive frequencies since $\Omega>0$ by definition.
Finally, this expression is converted straightforwardly
into the form (\ref{eq:Aphom_etar}) by using the relations
\begin{eqnarray}
  && \vartheta(\Omega)\varrho =
  \sqrt{\Omega^2-\omega_*^2}\,\varrho - \Omega\eta\,,
  \nonumber \\[10pt]
  && \frac{1}{\pi}
  \sqrt{\frac{\pi k_0^3}{\omega_*^2\varrho}}
  =
  \frac{1}{\sqrt{\pi}}
  \frac{(\Omega^2-\omega_*^2)^{3/4}}{\omega_*\sqrt{\varrho}}
  \,,
  \nonumber \\[10pt]
  && \tilde{f}(\Omega)
  = \tilde{\sigma}(\Omega,r_{\rm e})
  = |\tilde{\sigma}(\Omega,r_{\rm e})|\,
  e^{\iu \arg(\tilde{\sigma}(\Omega,r_{\rm e}))}\,,
\end{eqnarray}
where the last equation holds for a purely outgoing signal
emitted at $r_{\rm e}$.

\end{widetext}
\end{document}